\documentclass[preprint,12pt]{elsarticle}

\usepackage{amssymb}

\usepackage{lineno}
\usepackage{color,array}

\usepackage{graphicx}
\usepackage{makecell}
\usepackage{multirow}
\usepackage{amsmath}
\usepackage[autostyle]{csquotes}
\usepackage{amssymb}
\usepackage{caption}
\usepackage{subcaption}

\begin{document}

\begin{frontmatter}



\title{Understanding Information Spreading Mechanisms During COVID-19 Pandemic by Analyzing the Impact of Tweet Text and User Features for  Retweet Prediction}


\author{Pervaiz Iqbal Khan}

\address{German Research Center for Artificial Intllelligence (DFKI), Kaiserslautern, Germany and TU Kaiserslautern, Kaiserslautern, Germany (e-mail: pervaiz.khan@dfki.de)}
\author{Imran Razzak}
\address{Deakin University, Geelong, Austrailia (e-mail: imran.razzak@deakin.edu.au)}
\author{Andreas Dengel}
\address{German Research Center for Artificial Intllelligence (DFKI), Kaiserslautern, Germany and TU Kaiserslautern, Kaiserslautern, Germany (e-mail: andreas.dengel@dfki.de)}

\author{Sheraz Ahmed}
\address{German Research Center for Artificial Intelligence (DFKI), Kaiserslautern, Germany (e-mail: sheraz.ahmed@dfki.de)}
\begin{abstract}
COVID-19 has affected the world economy and the daily life routine of almost everyone. It has been a hot topic on social media platforms such as Twitter, Facebook, etc. These social media platforms enable users to share information with other users who can reshare this information, thus causing this information to spread. Twitter’s retweet functionality allows users to share the existing content with other users without altering the original content.  Analysis of social media platforms can help in detecting emergencies during pandemics that lead to taking preventive measures. One such type of analysis is predicting the number of retweets for a given COVID-19  related tweet. Recently, CIKM organized a retweet prediction challenge for COVID-19 tweets focusing on using numeric features only. However, our hypothesis is, tweet text may play a vital role in an accurate retweet prediction. In this paper, we combine numeric and text features for COVID-19 related retweet predictions. For this purpose, we propose two CNN and RNN based models and evaluate the performance of these models on a publicly available TweetsCOV19 dataset using seven different evaluation metrics. Our evaluation results show that combining tweet text with numeric features improves the performance of retweet prediction significantly.
\end{abstract}
 

\end{frontmatter}


\section{Introduction}
Coronavirus 2019 (COVID-19) originated from Wuhan, China, back in December 2019 has affected many countries across the world. On March 11, 2020, the World Health Organization (WHO) announced it a pandemic disease as it spread in 114 countries in the world \cite{ahmed2021deep}. On April 21, 2021, 143,589,434 cases of the virus have been reported worldwide, including 3,058,632 deaths \cite{worldo}. The total number of COVID-19 cases reported in USA alone are 32,536,470 including 25,105,535 recoveries, and 582,456 deaths as on April 21, 2021 \cite{worldo}. This virus has also affected the various public and private sectors such as tourism, airline and transportation, and private businesses that have hugely impacted the economy worldwide \cite{mansour2021sociodemographic}. To stop the spread of the virus, along with medical treatments, non-pharmaceutical interventions such as lock-downs, closing the educational institutions, local and international travel bans have been made \cite{shakibaei2021impact}. To make non-pharmaceutical interventions in a region, it is important to know the situation of the disease in that region. Analysis of social media platforms such as Twitter, Facebook, YouTube, etc., can be useful in knowing the COVID-19 situation in a region. For example, a tweet on Twitter related to COVID-19 may indicate how serious the situation of the virus is. \\
Twitter is an online social network service that allows users to share information with other users using short textual messages, called tweets. These tweets are a rich source of user communication. Their popularity has resulted in information propagation becoming a fundamental function of online social networks \cite{tang2015predicting,saeed2019event}. Twitter also provides a popular functionality known as retweeting a tweet that enables the user to share the content of an existing tweet from another user with his friends and followers without altering the original content. Retweeting is viewed as an atomic behavior and causes widespread information on the internet. Besides, a retweet is also an indicator that the user is interested in a particular tweet.  \\
Recently, the prediction of retweets has received significant attention. Retweet analysis has many applications, such as detecting and tracking the spread of fake news \cite{vosoughi2018spread}, and emergency management \cite{kogan2015think} in case of a pandemic. The more number of COVID-19 related retweets may indicate the prevalence of the disease. Hence predicting an accurate number of retweets is vital to analyze the situation of a virus. Information spread using retweet functionality depends on many factors such as the number of users mentioned in a tweet, number of friends, and followers of the user tweeting and retweeting the information. It also depends on the sentiments present in a tweet and the time of the tweet posting. For example, a tweet posted early morning or late at night is less likely to be retweeted, as most users may not be active on Twitter during this time. Tweet content such as text, image, or a video itself plays a role in retweeting a tweet. For example, a tweet \enquote{how’s self-quarantine going?}, with the highest number of retweets from the TweetsCOV19 dataset, contains a video that is the main factor of retweeting. However, the presence of videos and images in the tweet content poses additional challenges in retweet analysis. Most of the existing retweet prediction methods utilize information through modeling user preference such as user post history, user following relationship and user profile, etc. \cite{ma2019hot,daga2020prediction,lee2014will,wang2018retweet}. \\
Recently, CIKM-20 organized the COVID-19 retweet prediction challenge. The challenge focused on predicting the retweet frequency based on eleven features such as tweet ID, user name, timestamp, followers, friends, favorites, entities, mentions, URLs, hashtags, and sentiment. Although these features, especially sentiment, help predict the number of retweets, we believe that the content of a tweet also has an impact on retweet count. Fig. \ref{fig:ex} shows some of the retweeted tweets. To understand the spread of information related to the COVID-19 pandemic, in this work, we focus on the social media platform Twitter and try to predict the number of retweets for the given tweet. Along with numeric features present in the AnalytiCup competition dataset, we additionally utilize tweet text using deep learning methodologies for predicting retweet count. For this purpose, we propose two CNN and RNN based regression models. We perform experiments on these models in three different ways:
\begin{itemize}
    \item We feed these models with textual information only. To get statistical representation for tweet text, we initialize random embeddings that are learned during training.

    \item We feed these models with pre-processed numeric feature vectors only and remove the embedding layer.
    
    \item We reap the benefits of both textual input and pre-processed feature vectors. Here, we take both inputs at two branches, and after getting informative features from them, we concatenate and pass them to a dense layer for retweet probability prediction. 
\end{itemize}

 We evaluate the performance of the models on TweetsCOV19 dataset using seven different evaluation metrics. Experimental results show that combining tweet text with numeric features improve retweet prediction results significantly. The \textbf{key contributions} of this work are:

\begin{figure*}[!htb]
\centering
  \includegraphics[width=.8\textwidth]{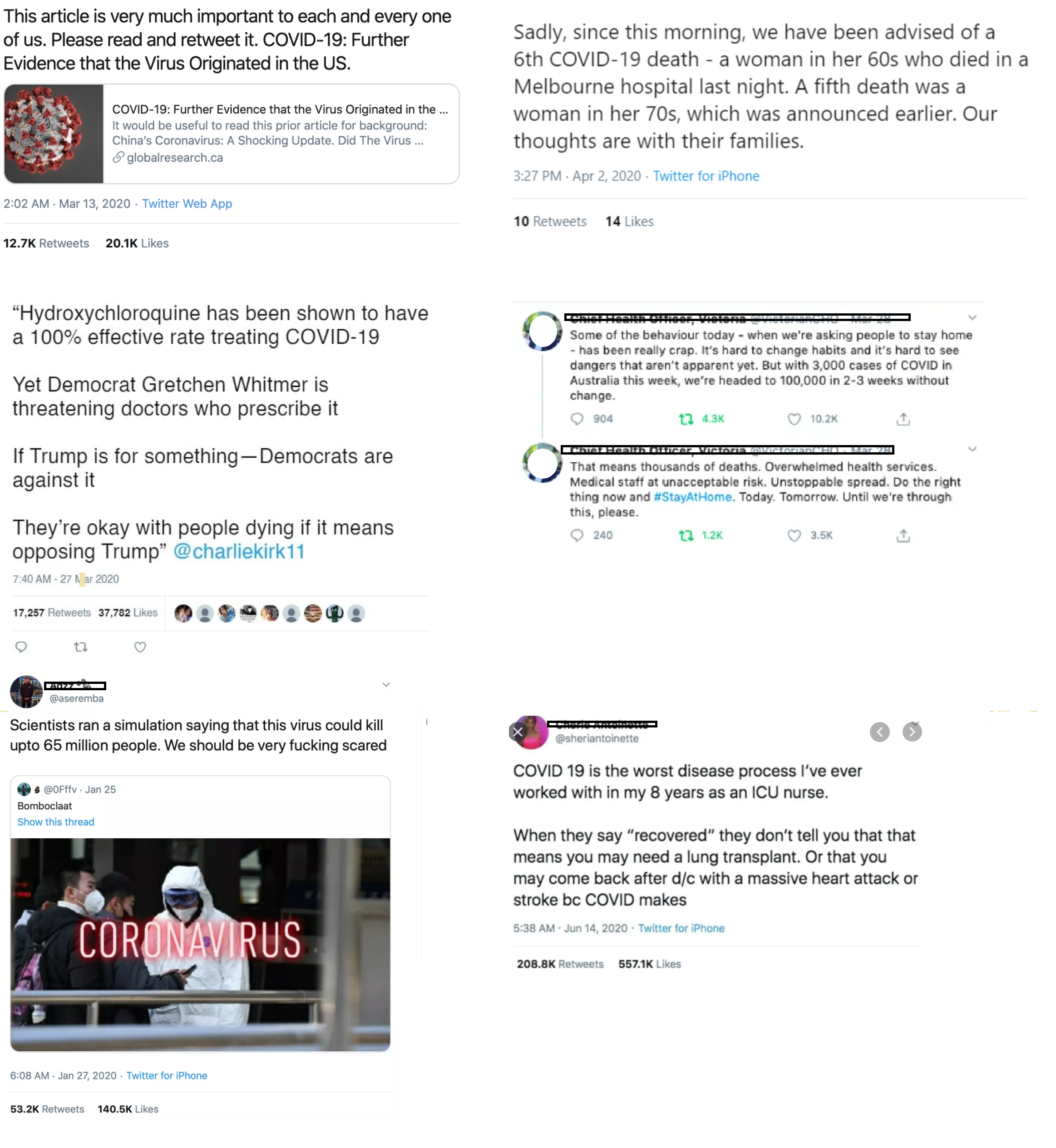}
  \caption{Motivation Example Tweets: Retweet number of different Tweet shows that Retweet count is significantly dependent on context of Tweet }
  \label{fig:ex}
\end{figure*}

\begin{itemize}
\item We propose a framework for predicting retweets for a COVID-19 related tweet based on CNN and RNN regression models.

    \item Unlike CIKM retweet challenge (AnalytiCup), we utilize both tweet text and retweet dataset features (6 features) to improve the prediction.

    \item As the understanding of COVID-19 related retweet behavior has many practical applications, we analyze the importance of features for the retweet prediction task.
    
  \item We conduct extensive experiments using various features contributing to retweets to demonstrate the effectiveness of tweet text for predicting retweet behavior. 
  
\end{itemize}

The structure of the rest of the paper is as follows: Section \ref{related-work} briefly reviews related work, and section \ref{def} presents the proposed retweet prediction framework. In section \ref{exp}, we provide the dataset and experimental details as well as present results and their discussion whereas, section \ref{conc}, concludes this paper.

\section{Related Work}\label{related-work}
This section briefly discusses the prior work done to predict the number of retweets for a given tweet. Zaman et al. \cite{zaman2010predicting} trained a probabilistic collaborative model called Matchbox \cite{stern2009matchbox} for the retweet prediction, developed originally for the prediction of the movie preference of a user. To collect data for the training and evaluation of the method, they crawled Twitter. They used tweeter and retweeter information such as the name of the user, the number of followers, etc., as well as tweet content itself as features for training the model. They introduced another feature called binary feedback and set it to 1 if the tweet was retweeted in the window of one hour of its original posting time and set to 0 otherwise. To evaluate the performance of the model, they used calibration plots as well as a negative log-score. However, they did not provide the training and testing of data distribution. \\

Firdaus et al. \cite{firdaus2016retweet} predicted the number of retweets for a given tweet based on the analysis of the user difference as an author and a retweeter. They considered the user personality and the topic of interest based on their tweet and retweet history. They used Big Five and their thirty lower level facet scores for each user as a personality measure, resulting in a 35-dimensional vector for each user based on their past tweets and retweets. They analyzed the frequent words used by users in their tweets and retweets. They also used the topic of interest as another feature based on their past tweets and retweets separately. For every user, they calculated the similarity score with the topic of interest of all other users. If the similarity score value was less than a threshold, this meant user interests were different as an author of the tweet and as a retweeter. Along with user profile generation based on their interests, they also created the profile for a given tweet based on its text and then compared the tweet text profile with the user profile to generate a new feature. Finally, machine learning classification was performed on features to classify whether the tweet would be retweeted by the users or not. Their results showed that considering user behavior differently as an author and retweeter outperformed the conventional methods.\\

Can et al. \cite{can2013predicting} predicted retweet count using visual cues in the tweet. They crawled Twitter for data collection and used the tweets for training and evaluating the model that only contained images. They experimented with structured-based features such as the number of friends, the number of followers, the number of favorites, and image-based features such as color histograms, GIST descriptors \cite{oliva2001modeling}, and object detectors \cite{li2010object}. For retweet prediction, they used machine learning models such as linear regression, Support Vector Machines (SVM) with Gaussian kernel, and random forest regression and used the Root Mean Squared Error (RMSE) as an evaluation metric. The experimental results showed that using image-based features on the random forest regression model achieved the best performance.  \\

Wang et al. \cite{wang2018retweet} studied the problem of retweet prediction using multimodal regression. They combined visual and textual data of tweets and the author’s social features such as number of friends, number of followers, etc., to predict the number of retweets for a given tweet. They used  Inception-ResNet CNN \cite{szegedy2016inception}  and LSTM-RNNs \cite{hochreiter1997long} to model the visual and textual features respectively. They used word embeddings specifically trained on tweet-style language, used as input to LSTM-RNNs. A joint embedding model was trained to learn the semantic relationship between tweet images and texts.  The learned visual, textual, and author’s social features were used as input to the Poisson regression model to predict the number of retweets. They trained and evaluated the model on an existing dataset MBI-1M dataset \cite{cappallo2015latent} and two other internal datasets containing tweets from 2015 and 2016. They used two evaluation metrics: Spearman ranking coefficient mean absolute percentage error (MAPE) for their model. It was evident from the results, naively combining visual, textual, and author’s social features did not improve repression model performance but via jointly embedding model. \\

As user exposure towards posting from followees can be used for retweet prediction, thus, Ma et al. \cite{ma2019hot} used hot topics discussed by followees using self-attentive model. In addition to this, authors have considered the user posting histories with external memory and utilize hierarchical attention modeling to construct users' interests. Zhang et al. presented a non-parametric statistical approach for predicting retweet behavior that combines textual, temporal, and structural information on a large number of microblogs and their social networks \cite{zhang2015retweet}. Microblog posts are aggregated to understand the topics of user concentration and the topic distribution of the cluster in the microblog is estimated.  As the topics may vary over time, thus weighted approach is used to increase the role of hot topics. Firdaus et al. \cite{firdaus2019topic} considered one of the influential and latent factors for retweet behavior and used topic-specific emotions as they may play a role in retweet prediction. Results showed that user profiles coupled with user emotion showed better performance in comparison to a user profile.  \\

Recently, CIKM organized a retweet prediction challenge (AnalytiCup) for predicting tweet popularity related to COVID-19 in terms of retweet frequency. Retweet prediction can be helpful during a crisis such as COVID-19. The challenge focused on eleven features, namely tweet ID, user name, timestamp, followers, friends, favorites, entities, mentions, URLs, hashtags, and sentiment, for retweet prediction. Although the tweet sentiment provides the tweet nature, we think that tweet text may also play a vital role in retweet prediction. In this work, we aim to analyze the impact of combining the tweet text with numeric features for the retweet prediction task and propose two CNN and RNN based regression models to perform experimentation.  
\section{Retweet Prediction Framework}\label{def}
Modeling the retweeting behavior is very important during the crisis time, and it has been an active area of research. Recently, the organized CIKM challenge AnalytiCup considers numeric features for the retweet prediction, however, we believe that the tweet text also plays a significant role in the retweet prediction. This work explores tweet text and numeric features behavior for the retweet prediction of a tweet. Let $T = \{ (t_i, y_i) \}_{i = 1}^{n}$ be the tweets, where $t_i$ and $y_i$ represent the $i$-th tweet and its number of retweets respectively and $n$ represents total tweets . Let $F = (f_1, f_2,f_3,f_4, f_5, f_6, f_7, f_8, f_9 )$ be the features from TweetsCOV19 dataset, and $F^{\prime}= (f_{text})$ be the textual features for every tweet $t_i$. Let $F^{*} = F + F^{\prime}$ represents the combined TweetsCOV19 dataset and textual features. The task is to predict $y_i$ for a given tweet $t_i$ using using features $F$, $F^{\prime}$, and $F^{*}$, and analyze the impact of these features.
Deep learning methodologies are achieving state-of-the-art performance for diverse natural language processing tasks such as text classification \cite{asim2020benchmark} \cite{asim2019two} \cite{asim2019robust}, sentiment analysis \cite{memood2020precisely}, information retrieval \cite{brants2003natural} and text summarization \cite{allahyari2017text}. Generally, deep learning methodologies are classified into two categories, namely Convolutional Neural Networks (CNNs) \cite{o2015introduction} and Recurrent Neural Networks (RNNs) \cite{sherstinsky2020fundamentals}. Initially, researchers believed that CNNs perform better in diverse tasks related to the computer vision (CV) domain whereas, RNNs achieve better performance in natural language processing (NLP) tasks. However, this thought is not valid as many researchers have concluded that CNN models also perform better on NLP tasks \cite{jacovi2018understanding}. To analyze the retweet behaviour, we propose two CNN and RNN based methods that take three different inputs such as numeric-only, text-only, and combined numeric and text features. Building-blocks of the proposed methods are given as follows: 
\begin{enumerate}
\item One-dimensional Convolution: The one-dimensional convolution (Conv-1D) operation computes the dot product between a weight vector $\textbf{w}\in\mathbb{R}^w$ and a vector of inputs $\textbf{x}\in\mathbb{R}^x$. Concretely, Conv-1D computes the dot product of weight vector $\textbf{w}$ with each $w$-th values in the input $\textbf{x}$ to obtain another vector $\textbf{y}$. The vector of weights $\textbf{w}$ is called filter or kernel and learned during the training of the network. In Conv-1D, the features residing at the margin of the input do not actively participate as compared to the features residing in the center. To prevent this, zero paddings is applied at the input and intermediate layers that ensure that all the weights in the filter reach the entire input sequence including the words at the margin.
\item k-Max Pooling: Given a sequence $\textbf{s}\in\mathbb{R}^s$ and a value k (where $s \ge k$), k-max pooling selects the subsequence $\textbf{s}^k_{max}$ highest values of $\textbf{s}$. The order of the values in $\textbf{s}^k_{max}$ corresponds to their original order in $\textbf{s}$. k-max pooling makes sure to select the k most active features from the input sequence $\textbf{s}$.

\item Non-linear Function: A non-linear activation function $g$ is applied element-wise to the input vector. Let a matrix $\textbf{M} \in \mathbb{R}^{f \times d}$, 
where $f$ is the number of filters and $d$ is the output dimensions from the pooling operation. Then the $i$-th activation vector $\textbf{a}_i$ for $i$-th filter is obtained as follows:

 \begin{align}
    \textbf{a}_i &= g\Bigg(\textbf{M}\begin{bmatrix}
           f_{i1} \\
           x_{i2} \\
           \vdots \\
           x_{id}
         \end{bmatrix}\Bigg)
  \end{align}
  
 \item Folding: After applying the 1D convolution, k-max pooling, and a non-linearity function to the input, a first-order feature map is obtained. These operations are repeated in each layer of the network to get more feature maps. Let $\textbf{F}^i$ denotes the features map in the $i$-th layer. Multiple features maps $\textbf{F}_1^{i}$, $\textbf{F}_2^{i}$,..., $\textbf{F}_n^{i}$ are computed in parallel in each layer, where $n$ denotes the number of filters in the $i$-th layer. These feature maps are independent of each other until they reach a fully connected layer. A simple method called folding sums every two rows in the feature map $\textbf{F}$. For a feature map of $m$ rows, folding returns $m/2$ rows i.e. half of the feature map rows. With the folding method, a feature map in the $i$-th layer now depends on two rows of the feature values in that layer.
 \item $\textbf{Recurrent Neural Networks:}$ At each time step $t$, an Recurrent Neural Network (RNN) takes an input $x_t$ and outputs a hidden state $h_t$. A hidden state is computed by using $X_t$ as well as the previous hidden state $x_{t-1}$.
\end{enumerate}

The CNN regressor uses Conv-1D, k-max pooling and non-linearity in the first layer whereas it uses Conv-1D, folding, k-max pooling and non-linearity in the second layer . The output of second layer is flattened and passed to the output layer. The RNN regressor on the other hand, consists of a simple RNN layer with 32 hidden units and non-linear activation function. The output of RNN layer is flattened and passed to the output layer for the prediction.

\begin{figure*}[!htb]
  \includegraphics[width=\textwidth]{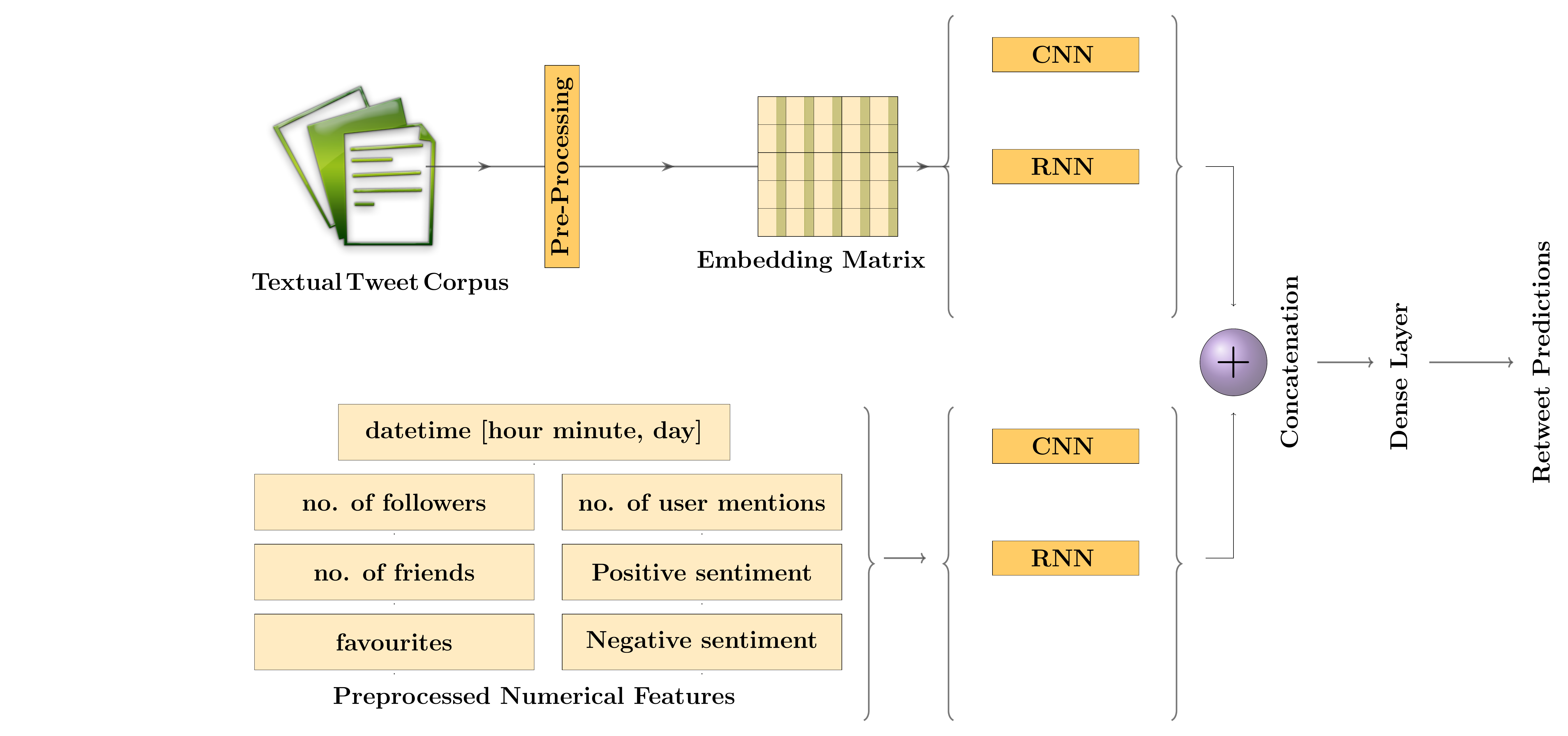}
  \caption{Block diagram of retweet prediction method}
  \label{fig:arch}
\end{figure*}

\section{Experiment}\label{exp}
Here, we present the dataset, evaluation metrics, and parameters used for training the model. Further, we provide the results of our experimentation and briefly discuss them.
\subsection{Dataset}
For our experimentation, we downloaded the TweetsCOV19 dataset containing tweets from the Twitter platform. The dataset contained a total of 8151524 tweets. Every tweet example had eleven features, namely Tweet\_id, Username, Timestamp, Number of Followers, Number of Friends, Number of Favorites, Entities, Sentiment, Mentions, Hashtags, URLs and a label \enquote{No. of Retweets}. As tweet text might contain important content, apart from these eleven features, we further crawled Twitter to obtain tweet text. However, at the time of crawling, not all the tweets were available for download, hence resulting in only 6955124 tweets with text. For our experimentation, we used a timestamp, the number of followers, number of friends, number of favorites, sentiment, mentions, and tweet text features instead of using all the features. We split the timestamp features into six features of timestamp month, timestamp week, timestamp day, timestamp hour, timestamp minute, timestamp day-of-week. The sentiment feature contained a score for positive sentiment ranging from 1 to 5 and a score for negative sentiment ranging from -1 to -5. Every tweet example contained positive and negative sentiment scores separated by a white-space character \enquote{ }. For example, sentiment value \enquote{3 -1} means that tweet has a positive sentiment score of 3 and a negative sentiment score of -1. Instead of directly using the sentiment feature, we split it into two features: Positive sentiment and negative sentiment. The user mention feature contained names of users mentioned in the tweet separated by white-space character \enquote{ }. We converted it into the total number of the user mentioned in the tweet. For example, if the user mention feature contained names of 3 persons, the feature value for this became 3. Table \ref{tab:stats} shows the top 5 most retweeted tweets from the dataset. As shown in the Table \ref{tab:stats}, tweet with the text \enquote{how’s self-quarantine going?} was the highest retweeted tweet with a total of 275529 retweets. This tweet also contained a video along with the tweet. The text was not available for the second tweet in the table at the time of download. The third most retweeted tweet also contained video along with the plain text on Twitter.  Instead of using all the tweets for our experimentation, we randomly took 60K tweets from the dataset and divide them into 40K, 10K, and 10K samples for train, evaluation, and test sets, respectively.

\begin{table}

 \begin{center}

     \resizebox{\textwidth}{!}{\begin{tabular}{ | c | c | c |}
    
       \hline
      \thead{Sr. No} & \thead{Tweet Text} & \thead{Retweets} \\
       \hline
   1 &  “how’s self quarantine going?” & 275529  \\
       \hline
      \hline
     2 &  Text not available  & 208891  \\
     \hline
      \hline
      3 & \makecell
 {This morning I tested positive for Covid 19. I feel ok, I have no\\ symptoms so far but have isolated since I found out about my possible\\ exposure to the virus.Stay home people and \\be pragmatic. I will keep you updated on how I’m doing. No panic.}  & 206181  \\
       \hline
      \hline
       4 & \makecell
{COVID-19 helping people realise that some meetings can be emails.}  & 189619  \\
       \hline
      \hline
      5 & \makecell
{I lost my Dad this morning to COVID-19. He was my rock, my best friend,\\and  my hero. He had virtually no symptoms and 48 hours later he was\\fighting for his life. I’m begging you guys from the bottom of my heart,\\  please stay inside and be safe.}  & 182990  \\
      \hline
     \end{tabular}}
    
   \end{center}
   \caption{\label{tab:stats}Retweet dataset containing top 5 most Retweeted Tweets.}
  \end{table}

\subsection{Evaluation Metrics}
To evaluate the performance of the models using numeric-only, text-only and combined numeric and text features, we used seven different evaluation metrics \cite{azhar2020evaluation}. Detail of each evaluation metric is given in the following sub-sections:

\subsubsection{Mean Absolute Error (MAE)}
Mean absolute error (MAE) calculates the expected or average enormity of errors in a set of predictions where directions are ignored \cite{mae}. Absolute differences between actual observations and predictions are computed and then we estimate the mean of these differences, where all differences have identical weight. Mathematical formula for MAE can be expressed as:
\begin{equation}
    MAE = \frac{\sum_{i=1}^{N}|(Y_i - \Bar{Y_i})|}{N}
\end{equation}
\
\subsubsection{Relative Mean Absolute Error (rMAE)}
Relative error helps to determine the magnitude of the absolute error in terms of the actual size of the measurement. Relative Mean Absolute Error (rMAE) is the normalized form of MAE. It is computed by dividing MAE by the mean of predictions.
Mathematical formula for rMAE can be expressed as:
\begin{equation}
    rMAE = \frac{MAE}{mean(\Bar{Y})}\times 100
\end{equation}
\
\subsubsection{Mean Bias Error (MBE)}
Mean Bias Error (MBE) calculates the expected or average enormity of errors in a set of predicted values. Differences are computed between actual observations and predictions and then we compute the mean of these differences, where all differences have same weight.
Mathematical formula of MBE can be expressed as:
\begin{equation}
    MBE = \frac{\sum_{i=1}^{N}(Y_i - \Bar{Y_i})}{N}
\end{equation}
\
\subsubsection{Relative Mean Bias Error (rMBE)}
Relative Mean Bias Error (rMBE) is the normalized form of MBE. It is computed by dividing MBE by the mean of predictions.
Mathematical formula for rMBE can be expressed as:
\begin{equation}
    rMBE = \frac{MBE}{mean(\Bar{Y})}\times 100
\end{equation}
\

\subsubsection{Root Mean Square Error (RMSE)}
Root Mean Square Error (RMSE) is a quadratic scoring technique that computes the average enormity of errors. First, we calculate the differences between actual observation and predictions then these differences are squared. We compute the mean of these squared differences and calculate the square root of this mean.
Mathematical formula of RMSE can be expressed as:
\begin{equation}
    RMSE = \sqrt{\frac{\sum_{i=1}^{N}(Y_i - \Bar{Y_i})^2}{N}}
\end{equation}
\
\subsubsection{Relative Root Mean Square Error (rRMSE)}
Relative Root Mean Square Error (rRMSE) is the normalized form of RMSE. It is computed by dividing RMSE by the mean of predictions.
Mathematical formula for rRMSE can be expressed as:

\begin{equation}
    rRMSE = \frac{RMSE}{mean(\Bar{Y})}\times 100
\end{equation}
\
\subsubsection{$R^2$ Score}
$R^2$ score is also termed as coefficient of determination. It shows the closeness between data and fitted regression boundary on data. In order to compute this metric, sum of squares of residuals (RSS) is divided by total sum of squares(TSS), and result is subtracted from 1.
Mathematical formula of $R^2$ can be expressed as:
\begin{equation}
    R^2 = 1 - \frac{RSS}{TSS}
\end{equation}

\subsection{Optimization Parameters}
Choosing an optimization approach for the deep learning model is an important aspect of the philosophy of deep learning since it can prove to be a move that saves time and resources by delivering outcomes in minutes instead of hours and in hours instead of days.

Adam \cite{kingma2019method} is one of the commonly used optimization algorithms. Instead of stochastic gradient descent, it can be used to change the network weights iteratively on training data. Adam was suggested by Diederik Kingma of OpenAI and Jimmy Ba of the University of Toronto in an ICLR (2015) paper. The paper was titled \enquote{Adam: A Stochastic Optimization Process} \cite{kingma2019method}.

The exponential moving average of the gradient and the squared gradient is calculated by Adam precisely. The decay rate of moving averages is regulated by two parameters, beta 1 and beta 2.

Following are the configuration parameters of Adam:
\begin{itemize}
    \item \textbf{alpha:} This parameter is the learning rate or step. It is the proportion with which weights get updated. Larger values of this parameter result in faster initial learning before the rate is updated. Learning rate is slowed down in training with a smaller value of this parameter.
    \item \textbf{beta1:} It is the exponential decay rate for first moment estimates.
    \item \textbf{beta2:} It is the exponential decay rate for second moment estimates. 
    \item \textbf{epsilon:} It is a very small number. This parameter is used to avoid division by zero during implementation.
\end{itemize}

\subsection{Training}
Proposed CNN regressor takes three types of inputs. For textual input, we first map all the unique words to number. Given the text-only input, replace each word with the number that generates a sequence $\textbf{s}$. Then, we take embedding $\textbf{e}\in\mathbb{R}^d$ for each token in the sequence $\textbf{s}$, where $d = 100$ in our settings. Thus it makes an embedding matrix $\textbf{E}^{dxl}$ where $l = 30$ is the maximum sequence length of our input. We apply zero-padding on both the side of size (49) that makes the size equal to $\textbf{E}^{100x128}$ and ensures that every token in the sequence participates actively during convolution operation. After that comes the Conv-1D, with  64 filters, followed by k-max-pooling with $d = 5$ and then non-linearity. Then we zero pad the out of the first layer, apply Conv-1D with the layer 64 filters, followed by a folding layer, k-max pooling and non-linearity. Then we flatten the output of this layer, and finally, a dense layer predicts the output. For the numeric-only input, the same configurations of the output are applied, except there is no embedding layer. For the combined numeric and text features, numeric and textual features are processed separately and combined the flattened outputs of both the features just before the final dense layer serving as a prediction layer.\\

\begin{figure*}[!htb]
     \centering
     \begin{subfigure}[b]{0.55\linewidth}
         \centering
         \includegraphics[width=\linewidth]{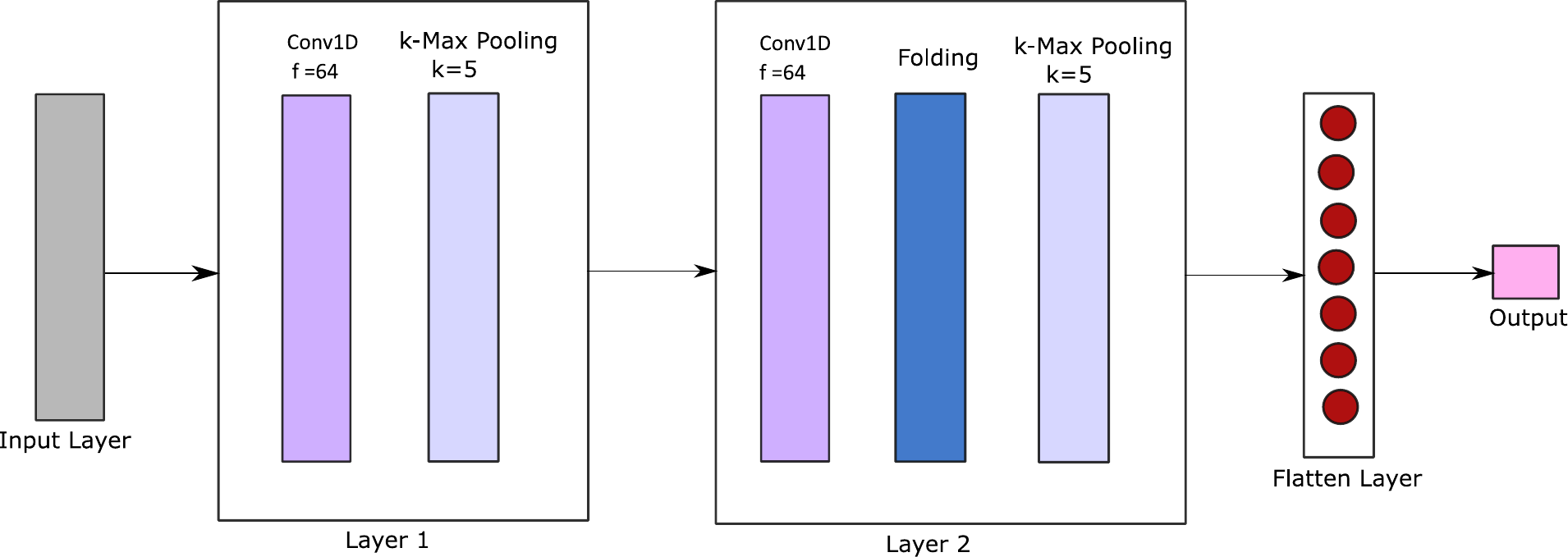}
         \caption{Architecture of CNN regressor}
         \label{fig:cnn}
     \end{subfigure}
     \hfill
     \begin{subfigure}[b]{0.44\linewidth}
         \centering
         \includegraphics[width=\linewidth]{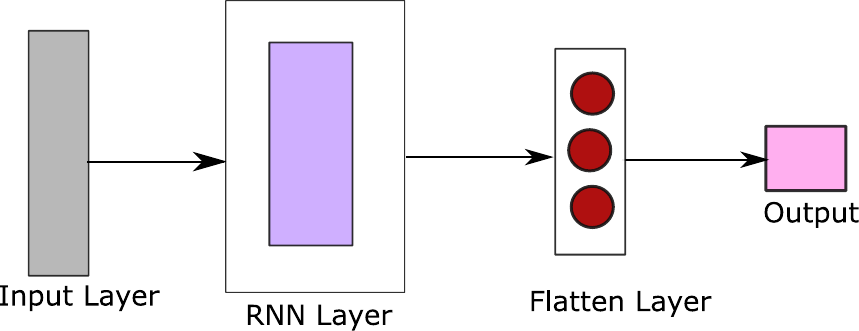}
         \caption{Architecture of RNN regressor}
         \label{fig:rnn}
     \end{subfigure}
     \caption{Architecture of the proposed regressors}
     \label{fig:regressors}

\end{figure*}

\begin{figure*}[!htb]
\centering
  \includegraphics[width=1.3\textwidth]{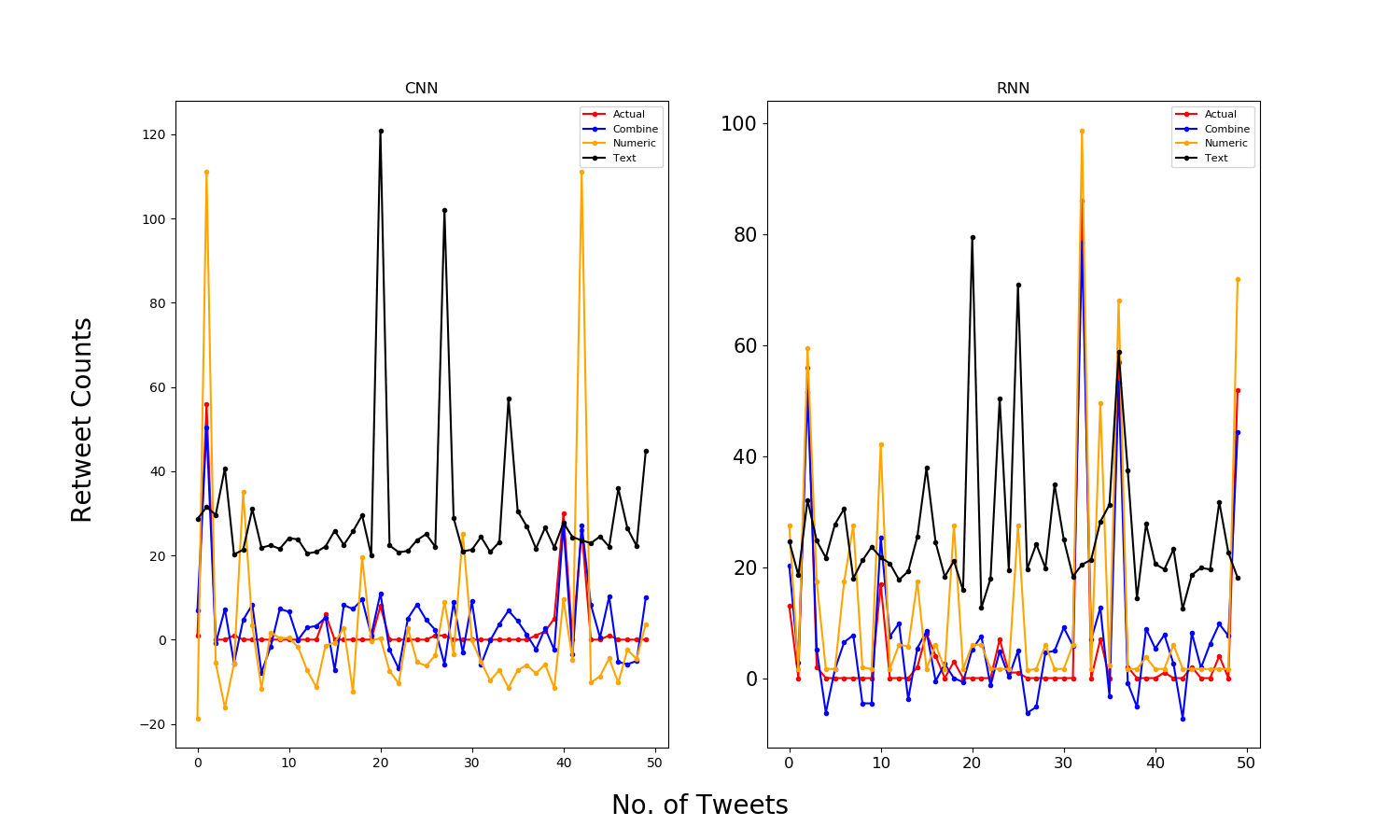}
  \caption{Plots showing actual and predicted retweets counts for the 50 randomly taken tweets from the test set.}
  \label{fig:plots}
\end{figure*}

We proposed an RNN based model to predict the number of retweets for a given tweet. This model consists of a simple RNN layer that has 32 hidden units and uses the $\tanh{}$ as an activation function. In the case of text-only features, it takes the input features, followed by an embedding layer similar to CNN regressor. The output of the embedding layer is passed to the RNN layer. The output of the RNN layer is flattened and passed to a dense layer for prediction. In the case of numeric-only features, there is no embedding layer and input features are directly passed to the RNN layer. For both the combined numeric and text features, both types of features are processed separately and combined before the prediction layer. Figure \ref{fig:arch} shows the block diagram of proposed method whereas Figure \ref{fig:regressors} presents the architecture of CNN and RNN regressors. In our experimentation, we set the value for the learning rate to be 0.001. We used Adam as an optimization function having default values of  beta 1 and beta 2 which were 0.9 and 0.999 respectively, and value for epsilon was 1e-07. We used a batch size of 64 whereas we trained both the models for 100 epochs.

\begin{table}[ht]
 \begin{center}
 
\resizebox{\textwidth}{!}{\begin{tabular}{|c|c|c|l|l|l|l|l|l|l|}
\hline
\multicolumn{1}{|l|}{\multirow{2}{*}{Sr. No}} & \multirow{2}{*}{Model}               & \multirow{2}{*}{Features} & \multicolumn{7}{c|}{Evaluation Metrics}                                                                                                                                                        \\ \cline{4-10} 
\multicolumn{1}{|l|}{}                        &                                      &                           & \multicolumn{1}{c|}{MAE} & \multicolumn{1}{c|}{rMAE} & \multicolumn{1}{c|}{MBE} & \multicolumn{1}{c|}{rMBE} & \multicolumn{1}{c|}{RMSE} & \multicolumn{1}{c|}{rRMSE} & \multicolumn{1}{c|}{$R^2$} \\ \hline
\multirow{3}{*}{1}                            & \multirow{3}{*}{CNN Regressor} & Numeric only              & 30.815                   & 76.35                     & -5.6342                  & 13.9597                   & 184.5064                  & 457.1499                   & 0.6466                  \\ \cline{3-10} 
                                              &                                      & Text only                 & 51.4912                  & 165.857                   & 3.6756                   & 11.8392                   & 310.2638                  & 999.3826                   & 0.0006                  \\ \cline{3-10} 
                                              &                                      & Combined            & \textbf{25.3887}         & \textbf{70.3753}          & \textbf{-1.3501}         & \textbf{3.7423}           & \textbf{157.4216}         & \textbf{436.361}           & \textbf{0.7427}         \\ \hline
\multirow{3}{*}{2}                            & \multirow{3}{*}{RNN Regressor} & Numeric only              & \textbf{37.2859}         & 148.3536                  & 9.5929                   & -38.1682                  & \textbf{299.4232}         & 1191.3481                  & \textbf{0.0692}         \\ \cline{3-10} 
                                              &                                      & Text only                 & 55.2734                  & 150.7142                  & -1.9532                  & 5.3259                    & 309.7459                  & \textbf{844.585}           & 0.004                   \\ \cline{3-10} 
                                              &                                      & Combined            & 48.9477                  & \textbf{142.6813}         & \textbf{0.4204}          & \textbf{-1.2255}          & 302.7707                  & 882.5696                   & 0.0483                  \\ \hline
\end{tabular}}
\end{center}
   \caption{\label{tab:results} Experimental results of CNN, and RNN based models for numeric only, text only and combined features.}
\end{table}

\subsection{Results and Discussion}
Here, we present the experimental results of models, we proposed to analyze the impact of tweet text with numeric features. We passed each of the CNN, and RNN models numeric-only, text-only, and combined numeric and text features. We evaluated the performance of both models using seven different metrics. It is evident from the results that the CNN-based model significantly performed better on all the seven evaluation metrics for combined numeric and text features. Numeric-only features performed better than text-only features. The model performed worst on text-only features.  The RNN-based model outperformed while using numeric-only features on the three evaluation metrics, namely mean absolute error, root mean squared error and $R^2$ score. It outperformed on only one metric for text-only features. This model outperformed while using combined numeric and text features on three evaluation metrics relative mean absolute error, mean bias error, and relative mean bias error.  Table \ref{tab:results} presents the experimental results of CNN and RNN models for numeric-only, text-only, and combined numeric and text features.

Based on the experimental evaluation, we have the following key observations:

\begin{itemize}
    \item Retweet prediction performance improved for the CNN-based model if numeric and text features are combined.
    \item For RNN-based models, text played an important role for retweet prediction either as a standalone feature or combining it with numeric features.
\end{itemize}
Based on these results, we can conclude that the tweet text plays a significant role in retweet prediction. Therefore, we should consider it for a retweet prediction instead of ignoring it.

Figure \ref{fig:plots} shows the predicted and actual retweet count on both the two models for 50 randomly taken tweets from the test set. Predicted retweet count is plotted for numeric-only, text-only, and combined numeric and text features. It is clear from the figure that the predicted retweets by combining numeric and text features are close to the actual number of retweets. Predicted retweets by using text-only features are far away from actual number of retweets.

\section{Conclusion}\label{conc}
In this paper, we predicted the number of retweets for a given COVID-19 related tweet using numeric features combined with tweet text. For this purpose, we proposed two CNN and RNN based models. We passed each of these models numeric-only, text-only, and combined numeric and text representation separately. We evaluated the performance of these models on the subset of the TweetsCOV19 (AnalytiCup) dataset. Results showed that CNN regressor achieved the highest $R^2$ of 0.7427, and the lowest MAE of 25.3887 for combined numeric and text features. On the other hand, the RNN regressor achieved the lowest rMAE of 142.6813, MBE of 0.4204, and rMBE of -1.2255 for combined numeric and text features. Experimental results showed that combining the numeric and text features improved retweet prediction as compared to individual numeric or text features.

\bibliographystyle{model1-num-names}
\bibliography{ref.bib}

\begin{thebibliography}{35}
\expandafter\ifx\csname natexlab\endcsname\relax\def\natexlab#1{#1}\fi
\providecommand{\bibinfo}[2]{#2}
\ifx\xfnm\relax \def\xfnm[#1]{\unskip,\space#1}\fi
\bibitem[{Ahmed et~al.(2021)Ahmed, Ahmad, Rodrigues, Jeon, and
  Din}]{ahmed2021deep}
\bibinfo{author}{I.~Ahmed}, \bibinfo{author}{M.~Ahmad}, \bibinfo{author}{J.~J.
  Rodrigues}, \bibinfo{author}{G.~Jeon}, \bibinfo{author}{S.~Din},
\newblock \bibinfo{title}{A deep learning-based social distance monitoring
  framework for covid-19},
\newblock \bibinfo{journal}{Sustainable Cities and Society}
  \bibinfo{volume}{65} (\bibinfo{year}{2021}) \bibinfo{pages}{102571}.
\bibitem[{Worldometer(2021)}]{worldo}
\bibinfo{author}{Worldometer},
\newblock \bibinfo{journal}{https://www.worldometers.info/coronavirus/}
  (\bibinfo{year}{2021}).
\bibitem[{Mansour et~al.(2021)Mansour, Al~Kindi, Al-Said, Al-Said, and
  Atkinson}]{mansour2021sociodemographic}
\bibinfo{author}{S.~Mansour}, \bibinfo{author}{A.~Al~Kindi},
  \bibinfo{author}{A.~Al-Said}, \bibinfo{author}{A.~Al-Said},
  \bibinfo{author}{P.~Atkinson},
\newblock \bibinfo{title}{Sociodemographic determinants of covid-19 incidence
  rates in oman: Geospatial modelling using multiscale geographically weighted
  regression (mgwr)},
\newblock \bibinfo{journal}{Sustainable cities and society}
  \bibinfo{volume}{65} (\bibinfo{year}{2021}) \bibinfo{pages}{102627}.
\bibitem[{Shakibaei et~al.(2021)Shakibaei, De~Jong, Alpk{\"o}kin, and
  Rashidi}]{shakibaei2021impact}
\bibinfo{author}{S.~Shakibaei}, \bibinfo{author}{G.~C. De~Jong},
  \bibinfo{author}{P.~Alpk{\"o}kin}, \bibinfo{author}{T.~H. Rashidi},
\newblock \bibinfo{title}{Impact of the covid-19 pandemic on travel behavior in
  istanbul: A panel data analysis},
\newblock \bibinfo{journal}{Sustainable cities and society}
  \bibinfo{volume}{65} (\bibinfo{year}{2021}) \bibinfo{pages}{102619}.
\bibitem[{Tang et~al.(2015)Tang, Miao, Quan, Tang, and
  Deng}]{tang2015predicting}
\bibinfo{author}{X.~Tang}, \bibinfo{author}{Q.~Miao},
  \bibinfo{author}{Y.~Quan}, \bibinfo{author}{J.~Tang},
  \bibinfo{author}{K.~Deng},
\newblock \bibinfo{title}{Predicting individual retweet behavior by user
  similarity: A multi-task learning approach},
\newblock \bibinfo{journal}{Knowledge-Based Systems} \bibinfo{volume}{89}
  (\bibinfo{year}{2015}) \bibinfo{pages}{681--688}.
\bibitem[{Saeed et~al.(2019)Saeed, Abbasi, Razzak, and Xu}]{saeed2019event}
\bibinfo{author}{Z.~Saeed}, \bibinfo{author}{R.~A. Abbasi},
  \bibinfo{author}{M.~I. Razzak}, \bibinfo{author}{G.~Xu},
\newblock \bibinfo{title}{Event detection in twitter stream using weighted
  dynamic heartbeat graph approach},
\newblock \bibinfo{journal}{arXiv preprint arXiv:1902.08522}
  (\bibinfo{year}{2019}).
\bibitem[{Vosoughi et~al.(2018)Vosoughi, Roy, and Aral}]{vosoughi2018spread}
\bibinfo{author}{S.~Vosoughi}, \bibinfo{author}{D.~Roy},
  \bibinfo{author}{S.~Aral},
\newblock \bibinfo{title}{The spread of true and false news online},
\newblock \bibinfo{journal}{Science} \bibinfo{volume}{359}
  (\bibinfo{year}{2018}) \bibinfo{pages}{1146--1151}.
\bibitem[{Kogan et~al.(2015)Kogan, Palen, and Anderson}]{kogan2015think}
\bibinfo{author}{M.~Kogan}, \bibinfo{author}{L.~Palen}, \bibinfo{author}{K.~M.
  Anderson},
\newblock \bibinfo{title}{Think local, retweet global: Retweeting by the
  geographically-vulnerable during hurricane sandy},
\newblock in: \bibinfo{booktitle}{Proceedings of the 18th ACM conference on
  computer supported cooperative work \& social computing}, pp.
  \bibinfo{pages}{981--993}.
\bibitem[{Ma et~al.(2019)Ma, Hu, Zhang, Huang, and Jiang}]{ma2019hot}
\bibinfo{author}{R.~Ma}, \bibinfo{author}{X.~Hu}, \bibinfo{author}{Q.~Zhang},
  \bibinfo{author}{X.~Huang}, \bibinfo{author}{Y.-G. Jiang},
\newblock \bibinfo{title}{Hot topic-aware retweet prediction with masked
  self-attentive model},
\newblock in: \bibinfo{booktitle}{Proceedings of the 42nd International ACM
  SIGIR Conference on Research and Development in Information Retrieval}, pp.
  \bibinfo{pages}{525--534}.
\bibitem[{Daga et~al.(2020)Daga, Gupta, Vardhan, and
  Mukherjee}]{daga2020prediction}
\bibinfo{author}{I.~Daga}, \bibinfo{author}{A.~Gupta},
  \bibinfo{author}{R.~Vardhan}, \bibinfo{author}{P.~Mukherjee},
\newblock \bibinfo{title}{Prediction of likes and retweets using text
  information retrieval},
\newblock \bibinfo{journal}{Procedia Computer Science} \bibinfo{volume}{168}
  (\bibinfo{year}{2020}) \bibinfo{pages}{123--128}.
\bibitem[{Lee et~al.(2014)Lee, Mahmud, Chen, Zhou, and Nichols}]{lee2014will}
\bibinfo{author}{K.~Lee}, \bibinfo{author}{J.~Mahmud},
  \bibinfo{author}{J.~Chen}, \bibinfo{author}{M.~Zhou},
  \bibinfo{author}{J.~Nichols},
\newblock \bibinfo{title}{Who will retweet this? automatically identifying and
  engaging strangers on twitter to spread information},
\newblock in: \bibinfo{booktitle}{Proceedings of the 19th international
  conference on Intelligent User Interfaces}, pp. \bibinfo{pages}{247--256}.
\bibitem[{Wang et~al.(2018)Wang, Bansal, and Frahm}]{wang2018retweet}
\bibinfo{author}{K.~Wang}, \bibinfo{author}{M.~Bansal}, \bibinfo{author}{J.-M.
  Frahm},
\newblock \bibinfo{title}{Retweet wars: Tweet popularity prediction via dynamic
  multimodal regression},
\newblock in: \bibinfo{booktitle}{2018 IEEE Winter Conference on Applications
  of Computer Vision (WACV)}, \bibinfo{organization}{IEEE}, pp.
  \bibinfo{pages}{1842--1851}.
\bibitem[{Zaman et~al.(2010)Zaman, Herbrich, Van~Gael, and
  Stern}]{zaman2010predicting}
\bibinfo{author}{T.~R. Zaman}, \bibinfo{author}{R.~Herbrich},
  \bibinfo{author}{J.~Van~Gael}, \bibinfo{author}{D.~Stern},
\newblock \bibinfo{title}{Predicting information spreading in twitter},
\newblock in: \bibinfo{booktitle}{Workshop on computational social science and
  the wisdom of crowds, nips}, volume \bibinfo{volume}{104},
  \bibinfo{organization}{Citeseer}, pp. \bibinfo{pages}{17599--601}.
\bibitem[{Stern et~al.(2009)Stern, Herbrich, and Graepel}]{stern2009matchbox}
\bibinfo{author}{D.~H. Stern}, \bibinfo{author}{R.~Herbrich},
  \bibinfo{author}{T.~Graepel},
\newblock \bibinfo{title}{Matchbox: large scale online bayesian
  recommendations},
\newblock in: \bibinfo{booktitle}{Proceedings of the 18th international
  conference on World wide web}, pp. \bibinfo{pages}{111--120}.
\bibitem[{Firdaus et~al.(2016)Firdaus, Ding, and
  Sadeghian}]{firdaus2016retweet}
\bibinfo{author}{S.~N. Firdaus}, \bibinfo{author}{C.~Ding},
  \bibinfo{author}{A.~Sadeghian},
\newblock \bibinfo{title}{Retweet prediction considering user's difference as
  an author and retweeter},
\newblock in: \bibinfo{booktitle}{2016 IEEE/ACM International Conference on
  Advances in Social Networks Analysis and Mining (ASONAM)},
  \bibinfo{organization}{IEEE}, pp. \bibinfo{pages}{852--859}.
\bibitem[{Can et~al.(2013)Can, Oktay, and Manmatha}]{can2013predicting}
\bibinfo{author}{E.~F. Can}, \bibinfo{author}{H.~Oktay},
  \bibinfo{author}{R.~Manmatha},
\newblock \bibinfo{title}{Predicting retweet count using visual cues},
\newblock in: \bibinfo{booktitle}{Proceedings of the 22nd ACM international
  conference on Information \& Knowledge Management}, pp.
  \bibinfo{pages}{1481--1484}.
\bibitem[{Oliva and Torralba(2001)}]{oliva2001modeling}
\bibinfo{author}{A.~Oliva}, \bibinfo{author}{A.~Torralba},
\newblock \bibinfo{title}{Modeling the shape of the scene: A holistic
  representation of the spatial envelope},
\newblock \bibinfo{journal}{International journal of computer vision}
  \bibinfo{volume}{42} (\bibinfo{year}{2001}) \bibinfo{pages}{145--175}.
\bibitem[{Li et~al.(2010)Li, Su, Fei-Fei, and Xing}]{li2010object}
\bibinfo{author}{L.-J. Li}, \bibinfo{author}{H.~Su},
  \bibinfo{author}{L.~Fei-Fei}, \bibinfo{author}{E.~P. Xing},
\newblock \bibinfo{title}{Object bank: A high-level image representation for
  scene classification \& semantic feature sparsification},
\newblock in: \bibinfo{booktitle}{Advances in neural information processing
  systems}, pp. \bibinfo{pages}{1378--1386}.
\bibitem[{Szegedy et~al.(2016)Szegedy, Ioffe, Vanhoucke, and
  Alemi}]{szegedy2016inception}
\bibinfo{author}{C.~Szegedy}, \bibinfo{author}{S.~Ioffe},
  \bibinfo{author}{V.~Vanhoucke}, \bibinfo{author}{A.~Alemi},
\newblock \bibinfo{title}{Inception-v4, inception-resnet and the impact of
  residual connections on learning},
\newblock \bibinfo{journal}{arXiv preprint arXiv:1602.07261}
  (\bibinfo{year}{2016}).
\bibitem[{Hochreiter and Schmidhuber(1997)}]{hochreiter1997long}
\bibinfo{author}{S.~Hochreiter}, \bibinfo{author}{J.~Schmidhuber},
\newblock \bibinfo{title}{Long short-term memory},
\newblock \bibinfo{journal}{Neural computation} \bibinfo{volume}{9}
  (\bibinfo{year}{1997}) \bibinfo{pages}{1735--1780}.
\bibitem[{Cappallo et~al.(2015)Cappallo, Mensink, and
  Snoek}]{cappallo2015latent}
\bibinfo{author}{S.~Cappallo}, \bibinfo{author}{T.~Mensink},
  \bibinfo{author}{C.~G. Snoek},
\newblock \bibinfo{title}{Latent factors of visual popularity prediction},
\newblock in: \bibinfo{booktitle}{Proceedings of the 5th ACM on International
  Conference on Multimedia Retrieval}, pp. \bibinfo{pages}{195--202}.
\bibitem[{Zhang et~al.(2015)Zhang, Gong, Guo, and Huang}]{zhang2015retweet}
\bibinfo{author}{Q.~Zhang}, \bibinfo{author}{Y.~Gong},
  \bibinfo{author}{Y.~Guo}, \bibinfo{author}{X.~Huang},
\newblock \bibinfo{title}{Retweet behavior prediction using hierarchical
  dirichlet process.},
\newblock in: \bibinfo{booktitle}{AAAI}, pp. \bibinfo{pages}{403--409}.
\bibitem[{Firdaus et~al.(2019)Firdaus, Ding, and Sadeghian}]{firdaus2019topic}
\bibinfo{author}{S.~N. Firdaus}, \bibinfo{author}{C.~Ding},
  \bibinfo{author}{A.~Sadeghian},
\newblock \bibinfo{title}{Topic specific emotion detection for retweet
  prediction},
\newblock \bibinfo{journal}{International Journal of Machine Learning and
  Cybernetics} \bibinfo{volume}{10} (\bibinfo{year}{2019})
  \bibinfo{pages}{2071--2083}.
\bibitem[{Asim et~al.(2020)Asim, Ghani, Ibrahim, Ahmad, Mahmood, and
  Dengel}]{asim2020benchmark}
\bibinfo{author}{M.~N. Asim}, \bibinfo{author}{M.~U. Ghani},
  \bibinfo{author}{M.~A. Ibrahim}, \bibinfo{author}{S.~Ahmad},
  \bibinfo{author}{W.~Mahmood}, \bibinfo{author}{A.~Dengel},
\newblock \bibinfo{title}{Benchmark performance of machine and deep learning
  based methodologies for urdu text document classification},
\newblock \bibinfo{journal}{arXiv preprint arXiv:2003.01345}
  (\bibinfo{year}{2020}).
\bibitem[{Asim et~al.(2019{\natexlab{a}})Asim, Khan, Malik, Razzaque, Dengel,
  and Ahmed}]{asim2019two}
\bibinfo{author}{M.~N. Asim}, \bibinfo{author}{M.~U.~G. Khan},
  \bibinfo{author}{M.~I. Malik}, \bibinfo{author}{K.~Razzaque},
  \bibinfo{author}{A.~Dengel}, \bibinfo{author}{S.~Ahmed},
\newblock \bibinfo{title}{Two stream deep network for document image
  classification},
\newblock in: \bibinfo{booktitle}{2019 International Conference on Document
  Analysis and Recognition (ICDAR)}, \bibinfo{organization}{IEEE}, pp.
  \bibinfo{pages}{1410--1416}.
\bibitem[{Asim et~al.(2019{\natexlab{b}})Asim, Khan, Malik, Dengel, and
  Ahmed}]{asim2019robust}
\bibinfo{author}{M.~N. Asim}, \bibinfo{author}{M.~U.~G. Khan},
  \bibinfo{author}{M.~I. Malik}, \bibinfo{author}{A.~Dengel},
  \bibinfo{author}{S.~Ahmed},
\newblock \bibinfo{title}{A robust hybrid approach for textual document
  classification},
\newblock in: \bibinfo{booktitle}{2019 International Conference on Document
  Analysis and Recognition (ICDAR)}, \bibinfo{organization}{IEEE}, pp.
  \bibinfo{pages}{1390--1396}.
\bibitem[{Memood et~al.(2020)Memood, Ghani, Ibrahim, Shehzadi, and
  Asim}]{memood2020precisely}
\bibinfo{author}{F.~Memood}, \bibinfo{author}{M.~U. Ghani},
  \bibinfo{author}{M.~A. Ibrahim}, \bibinfo{author}{R.~Shehzadi},
  \bibinfo{author}{M.~N. Asim},
\newblock \bibinfo{title}{A precisely xtreme-multi channel hybrid approach for
  roman urdu sentiment analysis},
\newblock \bibinfo{journal}{arXiv preprint arXiv:2003.05443}
  (\bibinfo{year}{2020}).
\bibitem[{Brants(2003)}]{brants2003natural}
\bibinfo{author}{T.~Brants},
\newblock \bibinfo{title}{Natural language processing in information
  retrieval.},
\newblock \bibinfo{journal}{CLIN} \bibinfo{volume}{111} (\bibinfo{year}{2003}).
\bibitem[{Allahyari et~al.(2017)Allahyari, Pouriyeh, Assefi, Safaei, Trippe,
  Gutierrez, and Kochut}]{allahyari2017text}
\bibinfo{author}{M.~Allahyari}, \bibinfo{author}{S.~Pouriyeh},
  \bibinfo{author}{M.~Assefi}, \bibinfo{author}{S.~Safaei},
  \bibinfo{author}{E.~D. Trippe}, \bibinfo{author}{J.~B. Gutierrez},
  \bibinfo{author}{K.~Kochut},
\newblock \bibinfo{title}{Text summarization techniques: a brief survey},
\newblock \bibinfo{journal}{arXiv preprint arXiv:1707.02268}
  (\bibinfo{year}{2017}).
\bibitem[{O'Shea and Nash(2015)}]{o2015introduction}
\bibinfo{author}{K.~O'Shea}, \bibinfo{author}{R.~Nash},
\newblock \bibinfo{title}{An introduction to convolutional neural networks},
\newblock \bibinfo{journal}{arXiv preprint arXiv:1511.08458}
  (\bibinfo{year}{2015}).
\bibitem[{Sherstinsky(2020)}]{sherstinsky2020fundamentals}
\bibinfo{author}{A.~Sherstinsky},
\newblock \bibinfo{title}{Fundamentals of recurrent neural network (rnn) and
  long short-term memory (lstm) network},
\newblock \bibinfo{journal}{Physica D: Nonlinear Phenomena}
  \bibinfo{volume}{404} (\bibinfo{year}{2020}) \bibinfo{pages}{132306}.
\bibitem[{Jacovi et~al.(2018)Jacovi, Shalom, and
  Goldberg}]{jacovi2018understanding}
\bibinfo{author}{A.~Jacovi}, \bibinfo{author}{O.~S. Shalom},
  \bibinfo{author}{Y.~Goldberg},
\newblock \bibinfo{title}{Understanding convolutional neural networks for text
  classification},
\newblock \bibinfo{journal}{arXiv preprint arXiv:1809.08037}
  (\bibinfo{year}{2018}).
\bibitem[{Azhar et~al.(2020)Azhar, Blanc, Asim, Imran, Hayat, Shahid, Ali
  et~al.}]{azhar2020evaluation}
\bibinfo{author}{M.~Azhar}, \bibinfo{author}{P.~Blanc},
  \bibinfo{author}{M.~Asim}, \bibinfo{author}{S.~Imran},
  \bibinfo{author}{N.~Hayat}, \bibinfo{author}{H.~Shahid},
  \bibinfo{author}{H.~Ali}, et~al.,
\newblock \bibinfo{title}{The evaluation of reanalysis and analysis products of
  solar radiation for sindh province, pakistan},
\newblock \bibinfo{journal}{Renewable Energy} \bibinfo{volume}{145}
  (\bibinfo{year}{2020}) \bibinfo{pages}{347--362}.
\bibitem[{Sammut and Webb(2010)}]{mae}
\bibinfo{editor}{C.~Sammut}, \bibinfo{editor}{G.~I. Webb} (Eds.),
  \bibinfo{title}{Mean Absolute Error}, \bibinfo{publisher}{Springer US},
  \bibinfo{address}{Boston, MA}, pp. \bibinfo{pages}{652--652}.
\bibitem[{Kingma and Ba(2019)}]{kingma2019method}
\bibinfo{author}{D.~P. Kingma}, \bibinfo{author}{J.~A. Ba},
\newblock \bibinfo{title}{A method for stochastic optimization. arxiv 2014},
\newblock \bibinfo{journal}{arXiv preprint arXiv:1412.6980}
  \bibinfo{volume}{434} (\bibinfo{year}{2019}).

\end{thebibliography}







\end{document}